\documentstyle[aasms4]{article}
\input epsf

\begin{document}
\title{\bf Modified cold-dark-matter models in light of 53W091, an old galaxy
at high $z$.}

\author{A. Kashlinsky\altaffilmark{1,2}}
\affil{$^1$NORDITA, Blegdamsvej 17, DK-2100 Copenhagen 0, Denmark}
\affil {$^2$Theoretical Astrophysics Center, Juliane Maries Vej 30, 
DK-2100 Copenhagen 0, Denmark}
\author{R. Jimenez\altaffilmark{3}}
\affil{$^3$Royal Observatory, Blackford Hill, Edinburgh EH9-3HJ, UK}
\authoremail{raul@roe.ac.uk, kash@nordita.dk}



\def\plotone#1{\centering \leavevmode
\epsfxsize=\columnwidth \epsfbox{#1}}

\def\wisk#1{\ifmmode{#1}\else{$#1$}\fi}

\def\lt     {\wisk{<}}
\def\gt     {\wisk{>}}
\def\le     {\wisk{_<\atop^=}}
\def\ge     {\wisk{_>\atop^=}}
\def\lsim   {\wisk{_<\atop^{\sim}}}
\def\gsim   {\wisk{_>\atop^{\sim}}}
\def\kms    {\wisk{{\rm ~km~s^{-1}}}}
\def\Lsun   {\wisk{{\rm L_\odot}}}
\def\Zsun   {\wisk{{\rm Z_\odot}}}
\def\Msun   {\wisk{{\rm M_\odot}}}
\def\um     {$\mu$m}
\def\sig    {\wisk{\sigma}}
\def\etal   {{\sl et~al.\ }}
\def\eg	    {{\it e.g.\ }}
\def\ie     {{\it i.e.\ }}
\def\bsl    {\wisk{\backslash}}
\def\by     {\wisk{\times}}
\def\half {\wisk{\frac{1}{2}}}
\def\third {\wisk{\frac{1}{3}}}

\begin{abstract}
The epoch of galaxy formation provides an important additional test 
of cosmological
theories. Cold-dark-matter (CDM) models with cosmological constant ($\Lambda$)
are designed to 
account for the observed excess power in galaxy distribution, but 
at the same time suppress the power on small scales pushing
galaxy formation to recent epochs. We point out that the recently
discovered high redshift galaxy, 53W091, with accurate age measurements 
(Dunlop et al 1996) provides a particularly important test of these models. 
In the flat $\Lambda$-dominated Universe, the redshift of formation of 53W091 
decreases with decreasing $\Omega$.
However, in the modified CDM models decreasing $\Omega$ suppresses the small
scale power in the density field and 
this effect turns out to be dominant. We estimate the mass of  the galaxy 
and show that it represents a very rare and unlikely
event in the density field of such models. 
Similar problems would occur in other 
modifications of the CDM cosmogonies.
\end{abstract}


\newpage

High $z$ objects constrain the small scale part of the
spectrum of the primordial mass density field that cannot be probed
directly by the large-scale structure or microwave background observations
(Efstathiou $\&$ Rees 1988; Kashlinsky $\&$ Jones 1991; Kashlinsky 1993;
Mo $\&$ Fukugita 1996).
It was argued that the excess power observed in galaxy surveys
(Maddox et al 1990, Picard 1991)
can be explained within inflationary framework by invoking the low
density
flat Universe dominated by cosmological constant, $\Lambda$$\equiv$$
3H_0^2 \lambda$, (Efstathiou et al 1990;
hereafter LCDM) or by mixing CDM
with massive neutrinos (Schaefer et al 1992; Klypin et al 1993; 
Taylor and Rowan-Robinson 1992; Davis et al 1992) and keeping $\Omega$=1
(hereafter MDM).  However such models
at the same time suppress power on
small scales requiring late galaxy formation. 
The recently discovered
high $z$ galaxy, 53W091, (Dunlop et al 1996; hereafter D96) with 
red colors and old stellar
population had to be formed at
high $z$ posing observational problem for such models.
In the LCDM models, the redshift of its formation decreases
with decreasing $\Omega h$,
but this also suppresses small scale power in the
density field and delays galaxy formation. We show that the
galaxy represents a very
rare and unlikely event in the density field of the modified CDM
models.

D96  have recently discovered an extremely red galaxy, 53W091,
at $z=$1.55.
Its red color is indicative of an old stellar population and its
blue apparent magnitude $V$=26
implies a large luminous mass. They have been able to identify
late-type stellar
absorption features in the spectrum of the galaxy which allow one to
determine the age.
The spectrum for 53W091 was obtained in the range 2000 to 3500 \AA,
where the main contribution to the integrated light comes from the
main sequence stars.
The latter assumption is reasonable
for galaxies with ages between 0.5 Gyr and  10 Gyr,
when the blue giants have
already become supernovae,
yet the horizontal branch is still red
($B-V$$>$0.4).
(The galaxy is unlikely to be younger than 0.5 Gyr since
in that case the spectrum
would have been very blue contrary to what is observed.) The absence of the
blue component also means that
the galaxy  can indeed be modeled with a single event of star formation.
Hence D96 built a series of synthetic
spectra at different ages with the main contribution coming from the
main sequence stars. This was done
using the set of stellar atmospheres models from Kurucz (1992) and a grid
of stellar interior models from Jimenez \& MacDonald (1997). 

The best fit was found for the age of $t_{age}$=3.5Gyr
(D96).  An independent evidence that the age of
53W091 cannot be less than 3 Gyr comes from transition
breaks in the spectrum.
The two of them, at 2600 and 2900 \AA ,
 were computed for different metallicities (1/5 \Zsun,
\Zsun, and 2 \Zsun);
the observed amplitudes showed that the breaks
cannot be reproduced if $t_{age}$$<$3Gyr (D96).
In order to check the robustness of the age determination, they computed
synthetic spectra for models with different metallicities. For the case
of 1/5$\Zsun$ the best fit is obtained for
$t_{age}$=4.0Gyr. Only if the metallicity of 53W091
were as high as $2\Zsun$ the best
fit is obtained for $t_{age}$=3.0Gyr, in good agreement with the
determination from the breaks method. The numbers
for $t_{age}$ were computed in D96 assuming 53W091 to be an
elliptical, i.e. assuming that the $\alpha$-nuclei elements
are enhanced with respect to the Sun with
the typical enhancement factors being [$\alpha$/Fe]=0.3-0.5.
Relaxing this assumption would change $t_{age}$ by only $4\%$
as detailed simulations of this effect  have shown 
(Salaris, Chieffi \& Straniero 1993).
We therefore follow D96 and adopt for the present discussion
the age uncertainty of the 53W091 galaxy to be no more than 0.5 Gyr with
the most likely value of $t_{age}$=3.5Gyr. Pushing the age to the lower
limit of the range, $t_{age}$=3Gyr, would at the same time require high
metallicity ($Z\geq$2 \Zsun) typical of nuclei
of ellipticals. (The galaxy was observed with an aperture of 4 '' making the
nucleus region unresolved).

In the context of inflationary cosmogonies, which generally require that the
Universe be flat (Kashlinsky, Tkachev, Frieman 1994),
these observations would force one to invoke a non-zero cosmological constant.
Fig.1 shows the redshift
$z_{gal}$ at which the galaxy 53W091 must have formed its first stars for
$\Omega$+$\lambda$=1 Universe. 
One can see that
the value of $z_{gal}$ decreases as \underline{both} $\Omega$ and $h$ decrease.
On the other hand, in the LCDM cosmogonies the
small-scale power is
also reduced as the \underline{product} $\Omega h$ decreases;
this would at the same time delay collapse of first galaxies until
progressively smaller $z$.

To quantify this we proceed in the manner outlined in Kashlinsky (1993). 
The power
spectrum of CDM models
is assumed to be initially of the Harrison-Zeldovich form. 
For tilted spectra (Cen et al 1992) our
conclusions will be similar; this case is briefly discussed later in the 
{\it letter}.
During radiation-dominated era only
superhorizon fluctuations
grow; this leads to the evolution/transfer in the original power spectrum.
In the LCDM models, the shape of the
power spectrum does not change after the
matter-radiation equality and until fluctuations turn non-linear and collapse.
In the MDM models
the evolution of the power spectrum shape stops after neutrinos become
non-relativistic.
We consider first the LCDM models.
The normalization scheme chosen below is independent of the redshift,$z_i$, 
chosen to normalize the density field, provided the latter
was still linear. Thus we take the power spectrum at some early epoch, $z_i$,
to be $P(k)$$\propto$$kT^2(k)$ with
the transfer function whose shape is fixed by the horizon scale at
the matter-radiation equality $\propto$$(\Omega h^2)^{-1}$. 
It is parametrized after Bardeen et al (1986):
$T(k)$=$\ln (1+a_0 k) [1+a_1 k + (a_2 k)^2 +(a_3 k)^3 +(a_4
k)^4]^{-1/4}/a_0 k$
with $a_0$=0.6$a_1$=0.14$a_2$=0.43$a_3$=0.35$a_4$=$2.34 (\Omega
h)^{-1}$Mpc. In the normalization we adopt lowering $\Omega h$ 
increases large-scale power but at the same time suppresses it
on small scales.

In order to compute the initial density field we normalize its rms to
$\Delta_8$, the amplitude that the fluctuations over comoving scale
$r_8$$\equiv$$8h^{-1}$Mpc had to have at $z_i$ in order to
reach unity  fluctuation in galaxy counts today.
I.e. any fluctuation that had amplitude of $\Delta_8$ at $z_i$ will grow
to the amplitude $\sigma_8$$\equiv$$1/b$ at $z$=0, where $b$ is the bias
factor fixed by matching
to the COBE results. With this normalization the rms fluctuation at $z_i$,
$\Delta(M)$, over the mass-scale  $M$ is given by:
$\Delta^2(M)$=$\Delta^2_8
\int_0^\infty k^3 T^2(k) W(kr) dk/\int_0^\infty k^3 T^2(k) W(kr_8) dk$
where $r(M)$=$1.25 (M/10^{12}\Msun)^\third (\Omega/h)^{-\third} h^{-1}$Mpc is
the
comoving scale
containing mass $M$. $W(x)$ is the window function chosen to filter the density
field. 
For brevity,
the results we present are for the top-hat case, but the numbers for Gaussian
$W(x)$ would not
differ substantially. Once parameters $\Omega,\Lambda$ are
specified the
value of $\Delta_8$ can be computed analytically.

In the expanding Universe density fluctuations grow until they reach a critical
overdensity at which
time they turn-around and collapse to form compact objects. In open Universe
with zero cosmological
constant the growth freezes at $1+z_{in}$$\simeq$$\Omega^{-1}$; if the Universe
is flat and has
non-zero cosmological constant this occurs at 
$1+z_{in}$$\simeq$$\Omega^{-\third}$.
We denote with $\delta_{col}(z)$ the
overdensity the fluctuation had to have at $z_i$ in order to
complete its collapse at
redshift $z$ as is required in order to form first stars and is also
implied by the total radius of 53W091 of only (8-15)$h^{-1}$Kpc.
As usual in the gravitational
clustering picture we assume that any region that had mass overdensity at
$z_i$ greater than
$\delta_{col}(z)$ will collapse by redshift $z$ (Press $\&$ Schechter
1974). Its value in the above
normalization scheme
can be characterized via $Q(z)\equiv \delta_{col}(z)/\Delta_8$ plotted
in Fig.1 of Kashlinsky (1993). 
For the case of $\Omega$+$\lambda$=1 it can be approximated
$Q(z)$$\simeq$$3 \Omega^{0.225} b^{2/3} (1+z)$ 
at $z$$>$$\Omega^{-\third}-1$;
this is valid for the values of $z_{gal}$ plotted in Fig.1.
The advantage of this normalization scheme is that the value of $Q(z)$ is
independent of the
initial epoch where the density field is fixed, provided
the latter is still in linear regime.
Thus a ``typical" structure that has collapsed by redshift $z$ must in these
models have \underline{total} mass $M$
such that $\Delta(M)/\Delta_8$$>$$Q(z)$. 
Conversely, if one knows the mass of the
galaxy that collapsed
to form stars at $z$ one can use the Press-Schechter 
prescription to estimate
the probability of that happening in a given model.

In order to compute the total \underline{luminous} mass of the galaxy we used
the integrated synthetic spectra from D96. 
The flux was then scaled
assuming the Miller-Scalo IMF until it matched the observed fluxes in all,
V,J,H,K, bands from D96. The effect
of changing the slope of the IMF on the total luminous mass is generally
small: $\sim$1\% when the IMF slope, $\alpha$, changes from 2.5 to 3.5.
We computed the total mass in stars in the galaxy for different cosmologies.
Table 1 summarizes the results assuming a single burst of star formation:
the first
three rows correspond to the Einstein-de Sitter Universe, $\Omega$=1, with
$h$=0.6 for the three values assumed for metallicity of the galaxy and for
the three values of $t_{age}$. The following three rows correspond to the
open Universe and $h=0.6$; the last three rows show variations with $H_0$ for
$\Omega$=1. Increasing
the metallicity, $Z$, of 53W091 decreases $M$, but the change is very small for
reasonable values of $Z$. Because of the increasing distances, 
for flat $\Omega$+$\lambda$=1 models the mass is even
higher. E.g. for the flat $\Lambda$-dominated Universe 
with $\Omega$=0.2 the total stellar mass for
53W091 is $1.8$$\times$$10^{12} M_{\rm \odot}$. The trends with $H_0$
and $\Omega$ are very similar to the $\Lambda$=0
case; for brevity we do not present the numbers here.
In order to account for different rates of star formation we
adopted the star formation prescription described in Chambers \& Charlot 
(1990) and computed a series of integrated
synthetic spectra with different values for the typical star formation rate
parameter $\tau$.
The change in the total star mass for the different star formation laws
is small, particularly in the likely case of $\tau$$<$2Gyr.
Larger $\tau$ lead to $Z$$\gg$$\Zsun$
and also give larger $M$: e.g. if $\tau$$\geq$3Gyr,the mass estimates
shown in Table 1 \underline{increase} by 37\%. The numbers vary
little with model or cosmological parameters and show that
53W091 has $\geq$$10^{12}\Msun$ in stars alone inside the aperture 
of 4''.

These values for the total luminous mass are consistent with an alternative
way to compute the mass assuming 53W091 to be an elliptical galaxy.
The aperture diameter of 4'' corresponds to the physical radius at $z$=1.55
of 8.5$h^{-1}$Kpc
if $\Omega$=1 and 14$h^{-1}$ Kpc if $\Omega$=0.1 and the Universe is flat.
These radii are typical of luminous sizes of giant ellipticals as
are the colours and stellar
populations of 53W091. If 53W091 is assumed to be an elliptical then
the fundamental plane (Dressler et al 1987) relations lead to the total luminous
mass of 10$^{12} M_{\rm \odot}$ (Peacock 1996) \underline{inside} the
aperture of 4'', in agreement with our estimates.

We therefore conclude that 53W091 is a 3.5 Gyr old galaxy at $z$=1.55, and
has luminous mass of $\geq$10$^{12} M_{\rm \odot}$. Its
age is not likely to be
below 3Gyr and its colours and other properties are consistent with it being
an elliptical. The total mass (including dark matter) is likely to be a factor
10 larger in which case our conclusions will be much stronger.

How likely is it that such a massive galaxy of \underline{at least}
$10^{12}\Msun$ has collapsed to  form stars within the total 
radius of $\simeq$(8-15)$h^{-1}$Kpc at the redshift plotted in Fig.1 for
the LCDM models? The quantity $\zeta$$\equiv$$Q(z_{gal})
\Delta_8/\Delta(M)$ is the
number of standard deviations the galaxy that collapsed at $z_{gal}$
on mass-scale $M$ would be in a given model. In models where the primordial
density field is assumed Gaussian, this quantity uniquely determines the
number density of objects of this (or greater) mass. $\zeta$ depends on the
bias parameter,  $\propto$$b^{2/3}$, which is determined by
fitting to the COBE/DMR data (Smoot et al. 1992, Bennett et al 1996). 
In general such normalization requires
$b$$\sim$2 for the LCDM models normalized to large-scale structure observed
in galaxy catalogs (Kashlinsky 1992, Efstathiou, Bond \& White 1992). 
The normalization procedure has been discussed in
great detail for the LCDM models in Stompor, Gorski \& Banday (1995) 
for the two-year COBE/DMR maps.
The bias factor implied by the normalization of the LCDM models to
the microwave background anisotropies is plotted there in Fig.3b vs
$\Omega h^2$. For the range of $\Omega h^2$ relevant
to this discussion this dependence of $b$  on $\Omega_{B}$
is weak for the values of $\Omega_B$ consistent
with the standard Big-Bang nucleosynthesis (SBBN). Increasing $\Omega_B$ above 
the SBBN values would lead to
further suppression of the small-scale power, whereas decreasing it would
not lead to any significant small-scale power increase. Note also, that 
normalising the spectrum to the four year COBE/DMR data (Bennett et al 1996)
would increase $b$ by $\simeq$10-15\%. The values of $\zeta$ for $M$=$10^{12}
\Msun$
are plotted versus $\Omega$ in Fig.2 for $\Omega$+$\lambda$=1 for various
values of $t_{age}$ and $h$. As in Fig.1 solid lines correspond to
$t_{age}$=3Gyr, dotted to 3.5 and dashes to 4 Gyr. Three types of each line
correspond to $h$=0.6, 0.8 and 1 going from bottom to top at large values
of $\Omega$ at the right end of the graph. The line for $t_{age}$=4Gyr and
$h$=1 lies above the box. As the figure shows, this galaxy must
represent an extremely rare fluctuation in the density field specified by the 
LCDM models.

For Gaussian density field the probability
of the region collapsing by the redshift $z_{gal}$ would be $P_M$=$\half {\rm
erfc}(\zeta/\sqrt{2})$. The Press-Schechter
prescription works best in the limit of high peaks, $\zeta \gg 1$, and allows
one to compute the number
density of the collapsed objects of mass greater than $M$ as
$n(>M)$=$2 \rho\int_M^\infty \frac{\partial P_M}{\partial M}d\ln M$; ``2" being
the ``fudge" factor traditionally introduced to normalize the counts of objects
correctly in the Press-Schechter formalism (Peacock \& Heavens 1990, 
Bond et al 1991). 
Fig.3a shows the expected comoving number density of such galaxies, $n(>M)$,
in units of ($h^{-1}$Gpc)$^{-3}$ vs the mass for
$t_{age}$=3.5Gyr. As was discussed
the luminous mass of 53W091 is \gt$10^{12}$\Msun and the total mass must be at
least a factor of 10 larger. The numbers for the comoving number density
$n(>M)$ in Fig.3 were computed for $\Omega$=0.1,0.2,0.3 and
$h$=1,0.8,0.6. 
The models not shown lie below
the box. Since $n(>M)$ is mainly determined by the corresponding $\zeta$ one
can easily combine Fig.3a and Fig.2 to estimate the expected number densities
at different $t_{age}$.

One can conclude from Fig.3a that within the framework of the 
LCDM models this object must be extremely rare in the Universe. There exists a
narrow range of parameters (\underline{total} mass of $10^{12}\Msun$, age of
\lt 3 Gyr, $\Omega$=0.1 and $h$$\geq$0.8) where one expects to find a few of
such
objects with each horizon, $R_{hor}$$\simeq$$6 h^{-1}$Gpc for $\Omega$=1, 
but for most
cases the number density of such objects is less than one per horizon volume.
Comparison with Fig.1 shows that the models that
allow the largest number of such galaxies are the ones that also place it
at the lowest $z_{gal}$. If indeed more such galaxies with old stellar
populations
are found at high $z$, this would be difficult to explain within the 
LCDM models. It is interesting to note in this context the
recent discoveries of several (proto)galaxy candidates at $z$$\simeq$4.5
(Hu \& McMahon 1996) and
$z$$>$6 (Lanzetta, Yahil \& Fernandez-Soto 1996). The first case corresponds 
to two L$y\alpha$-emitting objects
sufficiently far from the nearest quasar that they are identified as
(proto)galaxies
forming their first generation of stars (the spectra are consistent with
a galaxy whose integrated light is dominated by massive stars).
Lanzetta, Yahil \& Fernandez-Soto (1996) report the discovery of 6 objects at 
$z$$\simeq$6. These objects
have ultraviolet luminosities and sizes similar to nearby starbursting
galaxies,
and therefore they are identified with the first generation of stars in the
forming galaxy.

Thus it appears that galaxies were already not uncommon at $z$$\gt$5. Fig.3b
plots the expected
number density of such galaxies at $z$=5 vs their total mass for the 
LCDM models. 
The lines
that are not drawn would appear below the box. It is interesting to see that
the models that
predict the most abundance of objects like 53W091 at the same time predict very
few, if any, galaxies at $z$$\simeq$5 and vice versa.

One possibility to account for 53W091 within the framework of the 
LCDM models may be to supose that the stars 
formed in smaller collapsed regions that later merged into the observed
galaxy.
However, the galaxy looks like a normal old relaxed giant elliptical. The
physical
diameter size corresponding to the aperture of 4`` at
$z$=1.55 is 29, 20 and 17 $h^{-1}$Kpc for $(\Omega,\lambda)$=
(0.1,0.9),(0.1,0) and (1,0) respectively; this is typical of diameters of giant
ellipticals. It would
be hard to see how mergers could have led to the substructure dissipating its
energy to collapse to what appears a normal elliptical radius in such a short
time. The galaxy has very red colors, indicating no star formation for
the past
$t_{age}$\gt3Gyr, so this would further have to occur without significant star
formation
expected from merger events (Bender, Ziegler \& Bruzual 1996). 
Indeed,  if the galaxy had undergone mergers
3.5 Gyr ago, the following star formation events would have created a
significant blue
component in its spectrum contrary to what is observed.
Thus it is likely that for at least 3.5 Gyr  53W091 did not undergo any major
merging
events and its mass 3.5 Gyr ago had to be the same. Even invoking mergers would
not help much. At the relevant masses and the LCDM power spectra 
$\zeta$$\propto$$M^{0.1-0.15}$, so in order to decrease $\zeta$ appreciably
53W091 must have undergone a very large number of mergers in a short time.

The galaxy 53W091, its mass and its age are even more problematic for the
 MDM models
which require $\Omega=1$ and also suppress small scale power
(van Dalen \& Shaefer 1992) leading to late
galaxy formation (Kashlinsky 1993, Klypin et al 1993, Ma \& Bertschinger 1994). 
In this case the value of $z_{gal}$ is $>$5 if $h$\gt0.4 and $t_{age}$=3Gyr. If
$t_{age}$=3.5Gyr and $h>0.4$ the redshift $z_{gal}$$>$9. One can
see that even at the
values as low as $h$$\simeq$0.4 the redshift at which this galaxy must have
collapsed to form stars is
significantly larger than allowed by the MDM power spectra 
(Ma \& Bertschinger 1994).
A similar problem would also exist for the tilted CDM model suggested to
explain the excess power seen in galaxy catalogs (Cen et al 1992). Assuming the
primordial index of the power spectrum ($\propto$$k^n$) to be $n$=0.7, as
suggested in the tilted CDM model and taking the conservative values of
$10^{12}\Msun$ for the \underline{total} mass of 53W091 and
$t_{age}$\gt3 Gyr gives $\zeta$$>$$6.6 b^{2/3}$ 
if $h$=0.45 is assumed. Such models would require
$b\simeq 2$ (Cen et al 1992) making this galaxy a $\zeta$$>$10-sigma event. If
$n$=0.8 the value of $\zeta$$>$$6b^{2/3}$.
For $h$=0.45 and the standard CDM ($n$=1) which
requires $b$$\simeq$1 by the COBE/DMR data, but fails to account for the excess
power seen in galaxy catalogs (Maddox et al 1990), 
53W091 would correspond to
$\zeta$$>$$5.4 b^{2/3}$ standard deviations of the primordial density field
if the total mass were $10^{12} \Msun$.

The observational properties of 53W091 and the findings of galaxy candidates at
$z$\gt5 suggest that the Universe contains more collapsed galaxies at high
$z$ than the modified CDM models would predict. 
Such objects can be
most readily accommodated by assuming both 1) low $\Omega$
Universe and 2) the small-scale power in the primordial power spectrum
in excess of that given by simple inflationary models. E.g.  the
necessarily high redshift of galaxy formation can be produced
in string models (Mahonen, Hara \& Miyoshi 1995) or in the phenomenological
primeval baryon isocurvature model (Peebles 1987).

We thank John Peacock and James Dunlop for illuminating comments 
on 53W091. 
RJ acknowledges the hospitality and visiting support from TAC and
NORDITA during this project.  We are grateful to the support of the
NORDITA-Uppsala Astroparticle workshop where most of this work was done.
This work was supported in part by the Danish National Research Foundation through
its establishment of the Theoretical Astrophysics Center.

\newpage
\begin{table}
\begin{tabular}{|cccc|}
\hline
Age (Gyr) & 1/5 Z$_{\rm \odot}$ & Z$_{\rm \odot}$ & 2 Z$_{\rm \odot}$ \\
\hline\hline
3.0& 1.0 $\times$ 10$^{12}$ & 0.9 $\times$ 10$^{12}$ & 0.8 $\times$ 10$^{12}$
\\
3.5& 1.2 $\times$ 10$^{12}$ & 1.1 $\times$ 10$^{12}$ & 1.0 $\times$ 10$^{12}$
\\
4.0& 1.3 $\times$ 10$^{12}$ & 1.2 $\times$ 10$^{12}$ & 1.1 $\times$ 10$^{12}$
\\
\hline
\end{tabular}

\begin{tabular}{|cccc|}
Age (Gyr) & $\Omega=0.2$ & $\Omega=0.3$ & $\Omega=0.4$ \\
\hline\hline
3.0& 1.4 $\times$ 10$^{12}$ & 1.3 $\times$ 10$^{12}$ & 1.1 $\times$ 10$^{12}$
\\
3.5& 1.8 $\times$ 10$^{12}$ & 1.5 $\times$ 10$^{12}$ & 1.3 $\times$ 10$^{12}$
\\
4.0& 1.9 $\times$ 10$^{12}$ & 1.7 $\times$ 10$^{12}$ & 1.4 $\times$ 10$^{12}$
\\
\hline
\end{tabular}

\begin{tabular}{|cccc|}
Age (Gyr) & $H_0=60$ & $H_0=80$ & $H_0=100$ \\
\hline\hline
3.0& 1.1 $\times$ 10$^{12}$ & 1.4 $\times$ 10$^{12}$ & 1.7 $\times$ 10$^{12}$
\\
3.5& 1.5 $\times$ 10$^{12}$ & 1.9 $\times$ 10$^{12}$ & 2.3 $\times$ 10$^{12}$
\\
4.0& 1.7 $\times$ 10$^{12}$ & 2.0 $\times$ 10$^{12}$ & 2.4 $\times$ 10$^{12}$
\\
\hline
\end{tabular}
\caption{Total \underline{star} mass of 53W091 determined from
stellar models scaled to the observed flux. 
}
\end{table}
{\bf Figure captions}:\\
{\bf Fig.1}. The redshift, $z_{gal}$, when star formation in 53W091 was completed is
plotted vs $\Omega$ for $\Omega+\lambda=1$ Universe. Solid lines correspond
to $t_{age}$=3, dotted to 3.5 and dashes to 4Gyr. Three lines of each type
correspond to $h=0.6,0.8,1$ from bottom up.\\
{\bf Fig.2}. $\zeta$, the number of standard deviations of the primordial density
field 53W091 should be in the flat $\Lambda$-dominated CDM models is plotted
vs $\Omega$. $\zeta$ scales $\propto b^{-2/3}$ and is plotted for $b$
normalized
to the second year COBE/DMR maps. It
is plotted for the total mass of 53W091 of $10^{12}\Msun$. Same notation as
in Fig.1.Three types of each line
correspond to $h$=0.6, 0.8 and 1 going from bottom to top at large values
of $\Omega$ at the right end of the graph. The line for $t_{age}$=4Gyr and
$h$=1 lies above the box. \\
{\bf Fig.3}. (a) the predicted number density, $n(>M)$, of galaxies like 53W091 with
the redshift of formation plotted in Fig.1 is plotted vs their total
mass in units of
$10^{10} \Msun$. Solid lines correspond to $\Omega=0.1$:
they are for $h=1,0.8,0.6$ from top to bottom respectively. Dotted lines
correspond
to $\Omega=0.2$ and $h=0.8,0.6$ from bottom to top; the line for $h=1$ lies
below
the box. Dashes correspond to $\Omega=0.3$ and $h=0.6$. The lines not
drawn
fall below the box. (b) Same as in (a) only plotted at $z=5$. All lines go from
top to bottom in the decreasing order of $h=1,0.8,0.6$.\\
\clearpage
\begin{figure}
\centering
\leavevmode
\epsfxsize=1.0
\columnwidth
\epsfbox{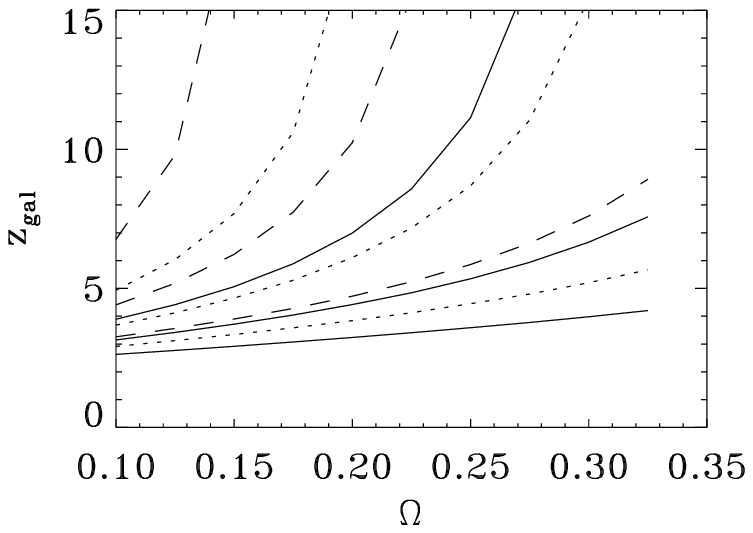}
\caption[]{}
\end{figure}
\clearpage
\begin{figure}
\centering
\leavevmode
\epsfxsize=1.0
\columnwidth
\epsfbox{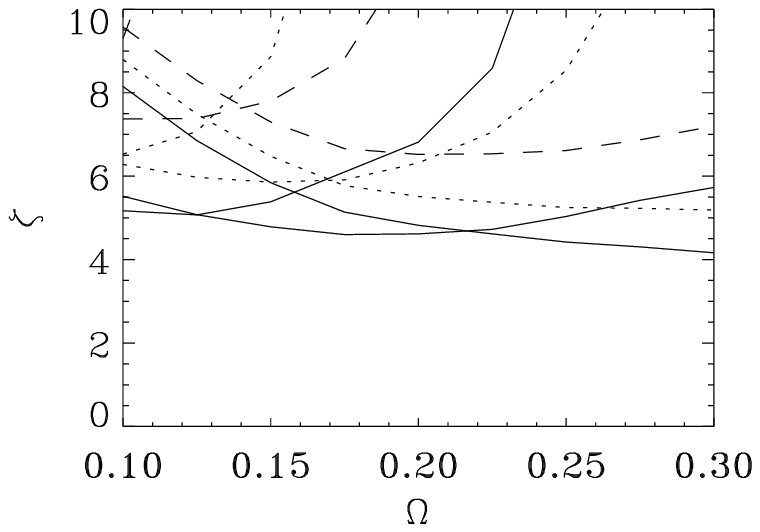}
\caption[]{}
\end{figure}
\clearpage
\begin{figure}
\centering
\leavevmode
\epsfxsize=1.0
\columnwidth
\epsfbox{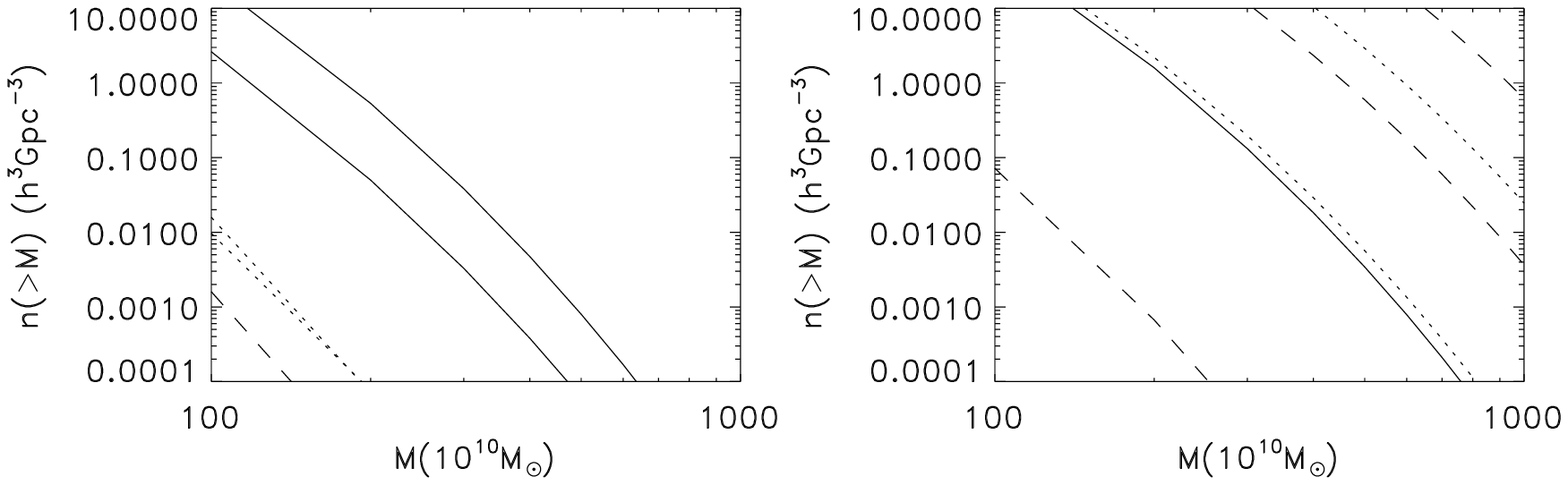}
\caption[]{}
\end{figure}


\begin{thebibliography}{}
\bibitem{} Bardeen, J.M., Bond, J.R., Kaiser, N., Szalay, A.S. 1986, Ap.J., {\bf
304}, 15.
\bibitem{} Bender, R., Ziegler, B. and Bruzual G. 1996, Ap.J., {\bf 463}, L51
\bibitem{} Bennett, C. \etal 1996, Ap.J.,{\bf 464}, L1.
\bibitem{} Bond, J.R., Kaiser, N., Cole, S., Efstathiou, G. 1991, Ap.J, {\bf
379}, 440.
\bibitem{} Cen, R. et al 1992, Ap.J., {\bf 393}, L11.
\bibitem{} Chambers, M. and Charlot, S., 1990, Ap.J., {\bf 348}, L1.
\bibitem{} Davis, M. et al. 1992, Nature, {\bf 359}, 393.
\bibitem{} Dressler, A. et al 1987, ApJ, {\
bf 313}, 42.
\bibitem{} Dunlop, J, Peacock, Spinrad, H., J., Dey, A., Jimenez, R. Stern, D. and
Windhorst, R. 1996, Nature, {\bf 381}, 581. (D96)
\bibitem{} Efstathiou, G. and Rees, M.J. 1988, MNRAS, {\bf 230}, 5P.
\bibitem{} Efstathiou, G., Sutherland, W. and Maddox, S. 1990, Nature, {\bf 348}, 705.
\bibitem{} Efstathiou, G., Bond, J.R., White, S.D.M. 1992,MNRAS,{\bf 258},1P.
\bibitem{} Hu, E.M. and McMahon, R.G. 1996, Nature, to appear.
\bibitem{} Jimenez and MacDonald 1997, {\it in preparation}.
\bibitem{} Kashlinsky, A. 1992, Ap.J.,{\bf 399},L1.
\bibitem{} Kashlinsky, A. 1993, Ap.J., {\bf 406}, L1.
\bibitem{} Kashlinsky, A. and Jones, B.J.T. 1991, Nature, {\bf 349}, 753.
\bibitem{} Kashlinsky, A., Tkachev, I. and Frieman, J. 1994, Phys.Rev., {\bf
73}, 1582.
\bibitem{} Klypin, A., Holtzman, J., Primack, J. and Regos, E. 1993, Ap.J., {\bf 416}, L1
\bibitem{} Kurucz, R. ATLAS9 Stellar Atmosphere Programs and 2km/s Grid CDROM Vol. 13
(Smithsonian Astrophysical Observatory, Cambridge, MA, 1992).
\bibitem{} Lanzetta, K.M., Yahil, A. and Fernandez-Soto, A. 1996, Nature, {\bf
381}, 759.
\bibitem{} Ma, C-P and Bertschinger, E. 1994, Ap.J., {\bf 434}, L5.
\bibitem{} Maddox, S., Efstathiou, G. Sutherland, W. and Loveday, J. et al
1990, MNRAS, {\bf
242}, 43P.
\bibitem{} Mahonen, P., Hara, T. and Miyoshi, S.J. 1995, Ap.J., {\bf 
454}, L81.
\bibitem{} Mo, H. and Fukugita, M. 1996, Ap.J., {\bf 467}, L9.
\bibitem{} Peacock, J.A. and Heavens, A.F. 1990, MNRAS, {\bf 243}, 133.
\bibitem{} Peacock, J., {\it private communication}.
\bibitem{} Peebles, P.J.E. 1987, Nature, {\bf 327}, 210.
\bibitem{} Picard, A. 1991, Ap.J., {\bf 368}, L7.
\bibitem{} Press and Schechter 1974, Ap.J., {\bf 187}, 425.
\bibitem{} Salaris, M., Chieffi, A., Straniero, O., 1993, ApJ, {\bf 414}, 580.
\bibitem{} Schaefer, R. et al 1992, Ap.J., {\bf 347}, 575.
\bibitem{} Smoot, G. \etal 1992, Ap.J., {\bf 396}, L1.
\bibitem{} Stompor, R., Gorski, K.M., Banday, A.J. 1995,MNRAS, {\bf 277}, 1225.
\bibitem{} Taylor, A.N. and Rowan-Robinson, M. 1992, Nature, {\bf 359}, 396.
\bibitem{} van Dalen, A. and Schaefer, R.K. 1992, Ap.J., {\bf 398}, 33.
\end{thebibliography}
\end{document}